\documentclass[%
 aip,
rsi,%
 amsmath,amssymb,
 reprint,%
onecolumn
]{revtex4-1}

\usepackage{graphicx}
\usepackage{dcolumn}
\usepackage{bm}

\def\d{\delta}

\def\v{\mathfrak{v}}

\begin{document}

\title{World-line instantons and the Schwinger effect as a WKB exact path integral}

\author{James Gordon}
\email{jbgordon@phas.ubc.ca}
\affiliation{ Department of Physics and Astronomy, University of British Columbia,  Vancouver, BC Canada V6T 1Z1}
\affiliation{Nordita, KTH Royal Institute of Technology and Stockholm University, Roslagstullsbacken 23,SE-106 91 Stockholm, Sweden}
\affiliation{Department of Physics and Astronomy, Uppsala University
SE-751 08 Uppsala, Sweden}
\author{Gordon W. Semenoff}
\email{gordonws@phas.ubc.ca}
\affiliation{ Department of Physics and Astronomy, University of British Columbia,  Vancouver, BC Canada V6T 1Z1}

\begin{abstract}
 A detailed study of the semiclassical expansion of the world line path integral for a charged
 relativistic particle in a constant external electric field is presented.   We show that the Schwinger formula
 for charged particle pair production is reproduced exactly by the semiclassical expansion around   
 classical instanton solutions when the leading order of fluctuations is taken into account.  We  
 prove that all corrections to this leading approximation vanish and that the WKB approximation to
 the world line path integral is exact.
\end{abstract}

\maketitle

\section{Introduction}

Schwinger's famous formula \cite{Schwinger:1951nm} for what is known as the ``Schwinger effect''  
gives the probability of the production of charged
particle-antiparticle pairs by a constant external electric field as 
\begin{align}P=1-e^{-\gamma V}\end{align} 
where, for spin zero particles,  the exponent is given by 
\begin{equation}\label{schwinger}
 \gamma=\sum_{n=1}^{\infty }\frac{\left(- 1\right)^{(n+1)} E^2}{8\pi ^3n^2} e^{-\pi
 m^2n/|E|}
\end{equation}
Here $m$  is the mass of the particles, $V$ is the space-time volume   and $E$ is the electric field. We
have absorbed a factor of the particle charge into $E$.

The result (\ref{schwinger}) is obtained by evaluating the vacuum persistence amplitude  in a theory 
with a charged massive scalar field exposed to an external electric field.  The phase in the persistence
amplitude, which normally contains the vacuum energy, obtains an imaginary part.   This gives a damping of
the amplitude which is attributed to  the production of charged particle-antiparticle pairs. 
The problem of finding the damping rate  
can be posed as that of evaluating the imaginary part of 
the world-line path integral for the relativistic particle,
\begin{align}\label{euclidean}
\gamma = -2\Im\frac{1}{V} \int_0^\infty\frac{dT}{T }\int [dx_\mu(\tau)]e^{-\int_0^1d\tau
\left[\frac{T}{4} \dot x_\mu(\tau)\dot x_\mu(\tau)+E
 x_1(\tau)\dot x_2(\tau)\right]
-\frac{m^2}{T}}
\end{align}
The integral is over periodic paths, $x_\mu(\tau+1)=x_\mu(\tau)$ and the space-time metric has Euclidean
signature. The variable $T$ as we use it here is the inverse of what is normally referred to as the ``Schwinger proper
time''.   Of course, a functional integral such as (\ref{euclidean}) must be defined with care.  In this paper, 
we will use zeta-function regularization in order to define the formally divergent infinite products and infinite summations
which are encountered in 
the course of computing (\ref{euclidean}). 
Basic formulae involving zeta functions are summarized in Appendix A.  
In   Appendix B, we shall give a detailed review of the derivation of this path integral formula from the
usual Feynman diagram representation of the vacuum persistence amplitude.  In particular, we demonstrate
that the path integral (\ref{euclidean}) with $E=0$ and with zeta function regularization reproduces 
the Feynman diagram expression for the vacuum energy in all of its details, including its normalization. 

The path integral in (\ref{euclidean}) can be
evaluated.  The real part can be presented as an integral over one variable and 
the imaginary part can be found to coincide with (\ref{schwinger}).  
The way that it is solved is to first perform the Gaussian integration 
over the position variables $x_\mu(\tau)$ in (\ref{euclidean}).  In this integration, 
the instability of the vacuum state of the system of charged particles when
a constant electric field is applied is reflected by the subtlety 
that the quadratic form in the Gaussian functional integral 
is not positive for all values of $T$. This is what allows this integral of a real function over real variables to have an imaginary part.   The integral is done by first assuming that $T$ is in a region

\vspace{0.5cm}
\noindent\textbf{Note added.~} The original version of this manuscript contained an error in the proof of semi-classical exactness (section \ref{exactness}), rendering the argument incomplete. In this version we correct the error and expand on the localization proof. This amendment is also included as a separate addendum/erratum to the published article \emph{J. Math. Phys.} \textbf{56} (2015) 022111. A proof based on supersymmetric localization is detailed in a separate note \cite{Gordon:2016ldj}.

\newpage 
\noindent where the Gaussian is stable, doing the 
Gaussian functional integral over $x_\mu(\tau)$ and then defining the result of the integral
for all values of $T$ by  analytic continuation.  The presence of values of $T$
where the path integral was unstable is then reflected as singularities in the remaining integration variable, $T$.
In this case, the singularities are simple poles on the real $T$-axis which
must be defined carefully to take causal boundary conditions into account.   
The imaginary part of the integral then comes from the  sum over the residues of the poles.  
This  yields the infinite series quoted
in (\ref{schwinger}) above.   This is straightforward.  
It leads to  the exponents in the individual terms in (\ref{schwinger}) 
and, with some care in normalizing the Gaussian functional integral involved,  
to the exact pre-factors in (\ref{schwinger}). 

There is another approach to computing the imaginary part of the 
path integral, which is less efficient, but it is often used as a starting point
for computations of the rate of particle production in the more general situation 
where the electric field is not constant \cite{Kim:2000un}-\cite{Dunne:2006ff}.   It has also been used
to discuss pair production in the context of AdS/CFT holography \cite{Semenoff:2011ng}.  This approach is a 
conventional semiclassical evaluation of the 
path integral.  It is generally good when the particle mass is large compared to other
dimensionful parameters.  In our case, the parameter which controls the semiclassical limit is the dimensionless ratio 
of the electric field strength to the mass squared, 
$\frac{E}{m^2}$,  which is small in  the ``weak field limit''.  
In this limit, we treat both $x_\mu(\tau)$ and $T$ as dynamical
variables and solve the classical equations of motion
which follow from the world-line action, 
 \begin{align}
S=\int_0^1 d\tau \left[ \frac{T}{4}\dot x_\mu(\tau)^2 +Ex_1(\tau)\dot x_2(\tau)+\frac{m^2}{T}\right]
\label{world-lineaction}
\end{align}
where $\dot x_\mu\equiv \tfrac{d}{d\tau}x_\mu(\tau)$.
The classical solutions, which we denote as   $x_{0\mu}(\tau)$ and $T_0$, are a saddle point of the path integral integrand.  We then compute
the integral by saddle point technique which amounts to changing integration variables as
\begin{align}
T&\to T_0+\delta T  \\
x_\mu(\tau)&\to x_{0\mu}(\tau)+\delta x_\mu(\tau)
\end{align}
and implementing perturbation theory in the  fluctuations $\delta T$ and $\delta x_\mu(\tau)$.  
This turns out to be an expansion in the parameter $\sqrt{\frac{E}{m^2}}$ and the expansion is valid in the regime where
this parameter is small.

There is a beautiful 
observation, due to Affleck, Alvarez and Manton \cite{Affleck:1981bma} that the classical solutions that are relevant to
the Schwinger process 
can be interpreted as instantons.  
The  $n$'th term in the summation in the Schwinger formula (\ref{schwinger}) can be interpreted as 
a $n$-instanton
amplitude in such a semi-classical computation of the path integral.  
They showed explicitly that the first, $n=1$ term in (\ref{schwinger}),
\begin{align}\label{one}
\gamma_1=\frac{ E^2}{8\pi ^3} e^{-\pi  m^2/|E|}
\end{align}
 is 
obtained exactly by such a semi-classical computation where they expand about a one-instanton solution
of the classical equations of motion for $x_\mu(\tau)$ and $T$.  The exponent in (\ref{one}) is the classical action
of the instanton.
The pre-factor is given by the Gaussian integral over fluctuations about the classical solution at the leading, quadratic order. 
It is interesting that, in the computation presented by Affleck, Alvarez and Manton,  
the integral is given exactly by what amounts to the leading orders of an approximation.  If it were an approximation, the small parameter
which suppresses corrections would be $\sqrt{\frac{E}{m^2}}$.  However, given that, in the leading orders they already obtained the exact 
result, 
as they noted, but did not demonstrate, higher
order perturbative corrections should then cancel exactly. This would mean that the computation has a much larger regime where it is valid,
in principle for all values of $\sqrt{\frac{E}{m^2}}$. 
 
The nature of the instanton is easy to understand.   In Euclidean space, a Minkowski space electric field behaves as a 
magnetic field.  In a magnetic field, the classical charged particle has a cyclotron orbit.  
The one-instanton solution is a single cyclotron orbit.  The  exponent of (\ref{one}) is simply the classical action 
of the world line theory evaluated on this orbit. The pre-factor in (\ref{one}) is given by evaluating the Gaussian
integral over the fluctuations about this classical solution.     
 
In this approach, the path integral gets an imaginary part due to the fact that the 
instantons in question are unstable solutions of the classical world-line theory.  
The unstable fluctuation turns out to be the fluctuation of the radius of
the cyclotron orbit.   The
Gaussian integral over the  fluctuations, including the fluctuation of the radius, 
then produces the square root of a determinant of a matrix
which has an odd number of negative eigenvalues, thus the factor of ``$i$''.   

As well as a single instanton that leads to (\ref{one}), there are an infinite series of multi-instanton 
classical solutions which are simply the multiple cyclotron orbits. 
In the following, we shall show that all of the higher terms in (\ref{schwinger}),
 \begin{align}\label{n}
\gamma_n=(-1)^{n+1}\frac{ E^2}{8\pi ^3n^2} e^{-\pi
 m^2n/|E|}~~,
\end{align} 
with $n=2,3,\ldots$ are produced by multi-instantons with higher wrapping number.
It is easy to see (and already well known) that the exponent of the $n$'th term as displayed in (\ref{n}) 
is the classical action of the $n$-instanton solution.  What we shall show is that the  
fluctuation integral produces the pre-factor of the exponential exactly.  This has the interesting
implication that the full, exact result is obtained in the semi-classical Gaussian approximation of the world-line path integral
where one sums over all of the classical solutions. 
Of course, the Gaussian approximation normally has corrections coming from expanding in the higher order non-Gaussian terms in the action, as
well as corrections from an expansion  about the classical solution of the terms which appear in the 
integration measure.  
Such terms are indeed at least formally present in this semiclassical expansion.  
What we shall find here, that the leading approximation produces  the exact result,
implies that the corrections must cancel.   We shall then give a
proof that this is indeed the case: all such corrections vanish.   The proof uses a simple scaling
argument together with a change of variables to localize the path integral on its semiclassical limit (see reference \cite{Gordon:2016ldj} for a fermionic symmetry-based argument).  This proof expands the range of validity of the semiclassical computation from the weak field limit to the strong field regime.   Whether this can help computations in less ideal problems, for 
example, where the electric field is not constant, is at this point an open question.

In section 2, we shall perform the semiclassical computation of the path integral in equation  (\ref{euclidean})
in the $n$-instanton sector.   We shall define the infinite products and sums which we encounter
using zeta function regularization.  
We show that, by careful treatment of the functional integration measure, 
the $n$'th term in the Schwinger formula (\ref{schwinger}), including the exact pre-factor, 
is obtained. 

In section 3, we examine higher order corrections beyond the leading order in the 
saddle point approximation. We find a proof that all
corrections beyond the integration of quadratic fluctuations must vanish.   The result is that, for computing the imaginary part of the vacuum persistence amplitude, the semiclassical limit of the world-line path integral with an external electric
field is exact.  

The definitions and values of the relevant zeta functions are summarized in
Appendix A.  A proof that the usual quantum scalar field theory vacuum energy derived
from the vacuum bubble Feynman diagram is identical to the world-line path integral defined 
using zeta function regularization is outlined in Appendix B.   In Appendix C we demonstrate the
semiclassical technique that we use on a simple example. In Appendix D we give an alternative, perturbative proof that all corrections to the semiclassical approximation vanish.

\section{Semiclassical evaluation of the world-line path integral} \label{semiclassical}

We shall begin with the case of a spinless charged particle of mass $m$ which is subject to a constant external
electric field.   Its vacuum energy is given by the world-line path integral (\ref{euclidean}).  The instability of the vacuum
to the production of on-shell particle-antiparticle pairs is reflected by the fact that the vacuum energy
has an imaginary part.  We shall compute this imaginary part in a semi-classical expansion about a classical
solution of the world-line theory.

 To begin, we shall first solve the classical equations of motion which are obtained by varying the world-line action
 by the dynamical variables $T$ and $x_\mu(\tau)$, 
\begin{align}
\frac{1}{4m^2}\int_0^1 \dot x^2&=\frac{1}{T^2}
~,~-\frac{T}{2}\ddot x_1 -E\dot x_2=0 ~,~
-\frac{T}{2}\ddot x_2  +E\dot x_1=0~,~
-\frac{T}{2}\ddot x_{3,4}&=0
\end{align}
with periodic boundary conditions, $x_\mu(\tau+1)=x_\mu(\tau)$. 
The solutions of these equations are
\begin{align}
x_1&= \frac{m}{E}\cos2\pi n\tau  ~,~
x_2= \frac{m}{E}\sin2\pi n\tau   ~,~
x_{3,4}=0  ~,~
T=\frac{E}{\pi n}
\end{align}
which we interpret as the $n$-instanton solution. 
Plugging these solutions into the  action (\ref{world-lineaction}) gives
$
S_{\rm cl.}=\frac{\pi n m^2}{E}
$, 
the same expression which appears in the exponents of the terms in (\ref{schwinger}).   

Now, we define the path   integration variables as the classical solutions plus
fluctuations, 
\begin{align}
x_1&= \frac{m}{E}\cos2\pi n\tau +\delta x_1 ~,~
x_2= \frac{m}{E}\sin2\pi n\tau  +\delta x_2~,~
x_{3,4}= \delta x_{3,4}~,~T=\frac{E}{\pi n} + \delta T
\end{align}
and we expand the action to quadratic order in the fluctuations.  We obtain
\begin{align}
S= \frac{\pi n m^2}{E}+ \frac{2m^2(\pi n)^3}{E^3}\frac{\delta T^2}{2} 
+\frac{m}{2E} (2\pi n)^2 \delta T\int d\tau (\cos(2\pi n\tau)\delta x_1+\sin(2\pi n\tau)\delta x_2)\nonumber \\ 
+  \frac{E}{4\pi n}\int d\tau \left[\delta\dot x^2 -4\pi n\delta x_1\delta\dot x_2\right]+\ldots 
\label{quadraticaction}
\end{align}
To proceed, we shall use the mode expansion
\begin{align}\label{modes}
\delta x_\mu(\tau)=x_\mu+\sum_{k=1}^\infty\left[ \sqrt{2}\cos(2\pi k\tau) a_{k\mu}+
\sqrt{2}\sin (2\pi k\tau) b_{k\mu}\right]
\end{align}
We first note that the action will not depend of the constant modes $x_\mu$.  These are space-time translation
zero modes. Their integration will result in the overall factor of the space-time volume $V$ in front of the functional
integral. 

When we substitute (\ref{modes}) into (\ref{quadraticaction}),   the action becomes
\begin{align}
S=& \frac{\pi n m^2}{E}
+ \frac{2m^2(\pi n)^3}{E^3} \frac{\delta T^2}{2} 
\nonumber \\ 
&+ \frac{ m(2\pi n)^2}{2  E}  \delta T  \left( \frac{ a_{n1}+b_{n2}  }{ \sqrt{2} }\right) 
+\frac{4\pi nE}{2 }\left(\frac {a_{n1}-b_{n2} }{\sqrt{2}}\right)^2 +
\frac{4\pi nE}{2 }\left(\frac{a_{n2}+b_{n1}}{\sqrt{2}}\right)^2        \nonumber \\ 
&+  \frac{E}{4\pi n  }\sum_{k=1,\neq n}^\infty (2\pi k)^2\left[(a_{k\mu}^2+b_{k\mu}^2)-\frac{2n}{k}(a_{k1}b_{k2}-a_{k2}b_{k1})\right]\nonumber \\
&+\frac{1}{4}\delta T\sum_k(2\pi k)^2[(a_k^\mu)^2+(b_k^\mu)^2]+\sum_{k=3}^\infty m^2
\left(\frac{\pi n}{ E}\right)^{k+1}(-\delta T)^k
\label{modeaction}
\end{align}
The last line of (\ref{modeaction}) contain terms of higher order than quadratic in the fluctuations.  
We have written them in this formula for future reference. 
To the leading order that we are studying in this section, they will be neglected.  

In the previous,  
quadratic terms in equation (\ref{modeaction}), 
we have separated the degrees of freedom $(a_{n1,2},b_{n1,2})$ which have the same frequency as the classical solution
and we have written them in the second line.
Note that the combination $a_{n2}-b_{n1}$ does not appear in the quadratic terms in the action -- this combination
is a zero mode.  The existence of the zero mode is due to a symmetry, the translation
invariance in $\tau$ of the action.  
The  integration measure, as we shall define it, is also invariant under translations of $\tau$.  
The world-line theory is thus $\tau$-translation invariant. However, the 
instanton solution depends on $\tau$ and it is not invariant.  The result is a zero mode in the fluctuations
about the solution. 

The way to handle the presence of a zero mode is by using the 
Faddeev-Popov trick to introduce a collective variable.  This technique
effectively substitutes $\delta ((a_{n2}-b_{n1})/\sqrt{2})$, accompanied by a Jacobian, into the integrand, 
and it multiplies the integral
by a factor of the volume of the symmetry group, $\int_0^1 d\tau =1$, in this case.  

The  introduction of a collective coordinate begins with inserting a factor of one into the path 
integral using  the identity
\begin{align}\label{faddeev-popov}
1=\frac{1}{\omega}\int_0^1 dt \delta(\chi(t))\left|\frac{d}{dt}\chi(t)\right|
\end{align}
where  the function $\chi(t)$ should be chosen so that the integration over the zero mode
becomes well-defined.  Here, 
$\omega$ is the number of solutions of $\chi(t)=0$ in the interval $t\in[0,1]$.  
We shall use the constraint
\begin{align}
&\chi(t)=\int_0^1d\tau(\sin(2\pi n\tau)  x_1(\tau-t)-\cos(2\pi n\tau)  x_2(\tau-t)))
\nonumber\\
&= \frac{1}{\sqrt{2}} \left[\left(\left[   \tfrac{m}{\sqrt{2}E}+a_{n1}\right] \sin(2\pi nt)  
+b_{n1}\cos(2\pi n t) \right) - \left( a_{n2}\cos(2\pi n t) -  \left[\tfrac{m}{\sqrt{2}E} +b_{n2}\right]\sin(2\pi nt)  \right) \right]
\end{align}
We shall later set $t=0$ by translating the time variable in the path integral.  The constraint reduces to
$ \chi(0)=\frac{1}{\sqrt{2}} \left[   
b_{n1}  -  a_{n2}  \right]$ which is what we need to constrain  the zero mode. 

As a function of $t$, $\chi(t)=0$ when
$$
\tan (2\pi n t) = \frac{a_{n2}-b_{n1}}{\frac{m}{\sqrt{2}E}+a_{n1}+\frac{m}{\sqrt{2}E}+b_{n2}}
$$
This equation is perodic in $t$ and it traverses $2n$ periods as $t$ varies from zero to one.
The fundamental domain (where it traverses one period) can be taken as 
$-\frac{1}{4n}<t<\frac{1}{4n}$ and has length $1/2n$. 
This fixes the constant in (\ref{faddeev-popov}) as $\omega=2n$. 

 The Jacobian evaluated on the constraint is 
\begin{align}
\frac{1}{\omega}\left|\frac{d}{dt}\chi(t) \right|_{\chi=0}= 
\frac{2\pi n}{\omega}\int_0^1 d\tau[\cos (2\pi n\tau) x_1(\tau)+\sin (2\pi n\tau) x_2(\tau)]=
 \left|
 \pi \frac{m}{E}+\pi  \frac{a_{1n}+b_{2n}}{\sqrt{2}}\right|
\end{align}

The net effect of this proceedure is the insertion of the delta function and measure factor
\begin{align}\label{fpfinal}
\delta \left(\frac{a_{n2}-b_{n1}}{\sqrt{2}}\right)~\left[~ \pi \frac{m}{E}
+\pi \frac{a_{1n}+b_{2n}}{\sqrt{2}}\right]
\end{align}
into the functional integral.  This suppresses the integration over the zero mode
and, in the leading order where we keep only the classical part of the Jacobian,  it inserts the factor
\begin{align}\label{11}
\pi \frac{m}{E}  
\end{align}
into the measure. 

We are now prepared to do the Gaussian integral.  The integration of the variables
$$\left[\delta T, \left( \frac{ a_{n1}+b_{n2}  }{\sqrt{2}}\right) 
,\left(\frac{a_{n1}-b_{n2}}{\sqrt{2}}\right),\left(\frac{a_{n2}+b_{n1}}{\sqrt{2}}\right)\right]$$   
gives the measure factor
\begin{align}
(2\pi)^2{\det}^{-\frac{1}{2}} \left[ 
\begin{matrix}
\frac{2m^2(\pi n)^3}{E^3} &  \frac{m(2\pi n)^2}{2E} &0&0 \cr 
\frac{m(2\pi n)^2}{2E} & 0 & 0 & 0 \cr
0 & 0& 4\pi nE &0 \cr
0 & 0 & 0 & 4\pi nE \cr 
\end{matrix}  \right]
  =\pm i (2\pi)^2\frac{  2E}{m(2\pi n)^2} \frac{1 }{ 4\pi nE }
\label{12}\end{align}
where the factor if $i$ arises from the fact that the determinant is 
negative, and the plus or minus reflects the fact that there is a choice
of sign when the square root is taken. (The mode with a negative eigenvalue
is called a ``tachyon''.)

Then, we can integrate over all of the other modes.  The result is the infinite product
\begin{align}
\prod_{k=1}^\infty \left(2\pi \frac{2\pi n}{E}\right)^2(2\pi k)^{-4}
\prod_{k=1,\neq n}^\infty
\left(2\pi \frac{2\pi n}{E}\right)^2(2\pi k)^{-4} \left(\det \left[\begin{matrix} 1 & 0 & 0 & -\tfrac{n}{k} \cr 0 & 1 & \tfrac{n}{k} & 0\cr
0 & \tfrac{n}{k} & 1 & 0 \cr -\tfrac{n}{k} & 0 & 0 & 1\cr \end{matrix} \right]\right)^{-\tfrac{1}{2}}
\nonumber \\
= 
\left(2\pi \frac{2\pi n}{E}\right)^{4\zeta(0)-2}(2\pi n)^{4}\prod_{k=1,\neq n}^\infty\frac{1}{1-\tfrac{n^2}{k^2}}
=\frac{E^4}{16\pi^4}\prod_{k=1,\neq n}^\infty\frac{1}{1-\tfrac{n^2}{k^2}}
\end{align}
In the above formula and in the following,  we define infinite products using zeta function regularization.  
Some of the conventions and the zeta functions that
are needed are reviewed in Appendix A. 

We find the identity
\begin{align}\label{13}
\prod_{k=1,\neq n}^\infty\frac{1}{1-\tfrac{n^2}{k^2}} = \lim_{\alpha\to n} \frac{(1-\frac{\alpha^2}{n^2})}{\prod_{k=1}^\infty (1-\frac{\alpha^2}{k^2})}
= \lim_{\alpha\to n}\frac{\pi\alpha(1-\frac{\alpha^2}{n^2})}{\sin\pi\alpha}=2(-1)^{n+1}
\end{align}
Gathering measure factors (\ref{11}), (\ref{12}) and (\ref{13}),
\begin{itemize}
\item{}The factor of $\frac{1}{T}$ in the integrand
$$
\frac {\pi n}{E} 
$$
\item{}The Faddeev-Popov determinant
$$
\pi \frac{m}{E}
$$
\item{}
The integral over the tachyon
$$
(\pm) i (2\pi)^2\frac{ 2 E}{m(2\pi n)^2} \frac{1 }{ 4\pi nE }
$$
\item{}The integral over all other modes
$$
\frac{E^4}{16\pi^4}\cdot 2(-1)^{n+1}
$$
\end{itemize}
we get the result
\begin{align}
\pi \frac{m}{E} \cdot \frac {\pi n}{E}  \cdot (\pm) i (2\pi)^2\frac{ 2 E}{m(2\pi n)^2} \frac{1 }{ 4\pi nE } \cdot 
\frac{E^4}{16\pi^4}\cdot 2(-1)^{n+1} =\pm i  \frac{E^2}{16\pi^3n^2}(-1)^{n+1}
\end{align}
With the appropriate choice of sign, and the factor of 2 from the formula (\ref{euclidean}), we
can see that we obtain, as the pre-factor of the exponential of the classical action, the 
factor $  \frac{E^2}{8\pi^3n^2}(-1)^{n+1}$ which matches the pre-factors of the exponential in each term in the summation (\ref{schwinger})
exactly.  The semiclassical integration has given us the exact result for the imaginary part of the integral in
the $n$-instanton sector, 
$
\gamma_n=(-1)^{n+1}\frac{ E^2}{8\pi ^3n^2} e^{-\pi
 m^2n/|E|}$. 
Summation of the instanton number results in the sum over $n$ which appears in the Schwinger
formula.  

It is interesting that we have produced the imaginary part of the functional integral exactly at this order
of what is putatively an approximate computation.  This means that all of the higher order corrections
to this approximation must cancel.  We shall explore this issue in the next section. 

\section{No more corrections} \label{exactness}

Now let us examine the corrections to the saddle point approximation which we
performed in the previous section.   Corrections arise  from the  expansion of the 
integrand   about the saddle-point.  If we expand the non-Gaussian parts of the 
integrand in a power series in the fluctuations, we can use the functional version of Wick's
theorem to compute the corrections.  Since we have already obtained the exact result in the
next-to-leading order of this expansion, we expect that the higher order corrections must
find a way to vanish.  In this Section, we shall prove that they indeed vanish. 

\setcounter{footnote}{0}
We consider the action (\ref{modeaction}) and we make the change of variables
\begin{align}
&x_\mu(  \tau)=\tilde  x_\mu(  \tilde \tau ) ~~,~~\tilde\tau = \tau\beta ~~,~~ \tilde x_\mu(\tilde\tau)=\tilde x_\mu(\tilde\tau+\beta)
\\ &T =\tilde  T/\sqrt{\beta}  
\end{align}
The path integral measure $[dx_\mu]$ is invariant under this change  of 
variables.
\footnote{To demonstrate this, we can show that the following Gaussian integral is independent of $\beta$:
$$
 \int [d x_\mu] e^{-\int_0^\beta d\tau  \frac{T}{4}\beta\dot x_\mu^2}~~,~~x_\mu(\tau+\beta)=x_\mu(\tau)
$$
Using the mode expansion (\ref{modeexpansion})
and  $[dx_\mu]=dx_\mu\prod_{n=1}^\infty da_{n\mu}db_{n\mu} $, 
\begin{align*}
 \int [ d x_\mu] e^{ -\int_0^\beta d\tau  \frac{T}{4}\beta\dot x_\mu^2}   &= \int dx_\mu 
 V\delta^D\left(\frac{x_\mu}{\sqrt{\beta}}\right) \prod_{n=1}^\infty da_{n\mu}db_{n\mu} 
 \exp(-\frac{T}{4\beta}\sum_{n=1}^\infty (2\pi n)^2(a_{n\mu}^2+b_{n\mu}^2) ) \\
 &= V   \beta^{D/2}\prod_{n=1}^\infty  
 \left[ \frac{4\pi\beta}{(2\pi n)^2 T}\right]^D = V\left[\frac{T}{4\pi}\right]^{D/2}
\end{align*}
where we have gauge fixed by inserting $1=\int dX^\mu\delta^D\left(X^\mu-
\frac{1}{\beta}\int_0^\beta d\tau x^\mu(\tau)\right)$. This results in the factor $ V\delta^D\left(\frac{x_\mu}{\sqrt{\beta}}\right) $
with $V$ the space-time volume.  We have also used zeta function regularization to define the infinite product.
The result does not depend on $\beta$. This is so in any dimension $D$.}
The scaling of $T$ cancels in the measure of the integral. 
The world-line action becomes (dropping the tildes)
\begin{align}
S= \int_0^\beta d\tau\left[ \sqrt{\beta}\frac{T}{4}\dot x^\mu(\tau)^2 +Ex^1(\tau)\dot x^2(\tau)\right]+\sqrt{\beta}\frac{m^2}{T}
\label{worldlineaction2}
\end{align}
The path integral cannot depend on the parameter $\beta$.
Moreover the limit where $\beta$ is large is the semiclassical limit.  In the following we shall take this limit with some care to show that it indeed projects   the full path integral to the semiclassical one which we computed in the previous section, where we only kept the classical and Gaussian terms in the action and the classical terms in the integration measure. Again, we shall expand the integration variables about the classical solution, 
\begin{align}
&x_1(\tau)= \frac{m}{E}\cos\frac{2\pi n\tau}{\beta} +\delta x_1(\tau) ~,~
x_2(\tau)= \frac{m}{E}\sin\frac{2\pi n\tau}{\beta} +  \delta x_2(\tau) \nonumber \\
&x_{3,4}(\tau)=\delta x_{3,4}(\tau) ~,~
T=\sqrt{\beta}\frac{E}{\pi n}+\delta T
\end{align}
with the fluctuations expanded as
\begin{align}\label{modeexpansion}
\delta x_\mu(\tau) =  \frac{x_\mu}{\sqrt{\beta}}+\sum_{k=1}^\infty \left[\sqrt{\frac{2}{\beta}}\cos\frac{2\pi k\tau}{\beta} a_{k\mu}
+\sqrt{\frac{2}{\beta}}\sin\frac{2\pi k\tau}{\beta} b_{k\mu}  \right]
\end{align}
The measure in the path integral is now
{\small\begin{align}
\frac{ d\delta T}{\sqrt{\beta}\frac{E}{\pi n}+\delta T}  V\delta\left(\frac{x_\mu}{\sqrt{\beta}}\right) 
\delta\left(\frac{b_{n1}-a_{n2}}{\sqrt{2\beta}}\right)\left[ \frac{\pi m}{E}+\frac{\pi}{\sqrt{\beta}}
\left(\frac{a_{n1}+b_{n2}}{\sqrt{2}}\right)\right]~\prod_{\mu=1}^4dx_\mu\prod_{k=1}^\infty da_{k\mu}db_{k\mu}
\label{measure}\end{align}}
The action becomes 
\begin{align}
S=& \frac{\pi n m^2}{E}
+ \frac{2m^2(\pi n)^3 }{\beta E^3} \frac{\delta T^2}{2} 
\nonumber \\ 
&+ \frac{ m(2\pi n)^2}{2 \beta E}  \delta T  \left( \frac{ a_{n1}+b_{n2}  }{ \sqrt{2} }\right) 
+\frac{4\pi nE}{2\beta }\left(\frac {a_{n1}-b_{n2} }{\sqrt{2}}\right)^2 +
\frac{4\pi nE}{2 \beta}\left(\frac{a_{n2}+b_{n1}}{\sqrt{2}}\right)^2        \nonumber \\ 
&+  \frac{E}{4\pi n \beta }\sum_{k=1,\neq n}^\infty (2\pi k)^2\left[(a_{k\mu}^2+b_{k\mu}^2)-\frac{2n}{k}(a_{k1}b_{k2}-a_{k2}b_{k1})\right]\nonumber \\
&+\frac{1}{4\beta^{\frac{3}{2}}}\delta T\sum_{k=1}^\infty(2\pi k)^2[(a_k^\mu)^2+(b_k^\mu)^2]+\sum_{k=3}^\infty m^2\frac{1}{\beta^{\frac{k}{2}}}
\left(\frac{\pi n}{ E}\right)^{k+1}(-\delta T)^k \label{actionbeta}
\end{align}
Now, in order to make the quadratic terms $\beta$-independent, we rescale
\begin{align}
\delta x^\mu(\tau)&\to \sqrt{\beta}\delta x^\mu(\tau)  
\end{align}
The Jacobian for this transformation is $\left(\sqrt{\beta}\right)^{4+8\zeta(0)} = 1$ where we have used $\zeta(0)=-1/2$. The integration measure becomes
$$
\frac{ d\delta T}{\sqrt{\beta} \frac{E}{\pi n}+\delta T}  V\delta\left( x_\mu \right) 
\delta\left(\frac{b_{n1}-a_{n2}}{\sqrt{2}}\right)\left[ \frac{\pi m}{E}+ \pi
\left(\frac{a_{n1}+b_{n2}}{\sqrt{2}}\right)\right]~\prod_{\mu=1}^4dx_\mu\prod_{k=1}^\infty da_{k\mu}db_{k\mu}
$$
and the action is 
\begin{align}
S=& \frac{\pi n m^2}{E}
+ \frac{2m^2(\pi n)^3 }{\beta E^3} \frac{\delta T^2}{2} 
\nonumber \\ 
&+ \frac{ m(2\pi n)^2}{2 E \sqrt{\beta}}  \delta T  \left( \frac{ a_{n1}+b_{n2}  }{ \sqrt{2} }\right) 
+\frac{4\pi nE}{2 }\left(\frac {a_{n1}-b_{n2} }{\sqrt{2}}\right)^2 +
\frac{4\pi nE}{2  }\left(\frac{a_{n2}+b_{n1}}{\sqrt{2}}\right)^2        \nonumber \\ 
&+  \frac{E}{4\pi n  }\sum_{k=1,\neq n}^\infty (2\pi k)^2\left[(a_{k\mu}^2+b_{k\mu}^2)-\frac{2n}{k}(a_{k1}b_{k2}-a_{k2}b_{k1})\right]\nonumber \\
&+\frac{1}{4\beta^{\frac{1}{2}}}\delta T\sum_{k=1}^\infty(2\pi k)^2[(a_k^\mu)^2+(b_k^\mu)^2]+\sum_{k=3}^\infty m^2\frac{1}{\beta^{\frac{k}{2}}}
\left(\frac{\pi n}{ E}\right)^{k+1}(-\delta T)^k
\end{align}
As it stands, we cannot directly set $\beta$ to infinity; in this limit the measure diverges, as does the integral over $\left( \frac{ a_{n1}+b_{n2}  }{ \sqrt{2} }\right)$ and $\d T$. Therefore we further rescale the single mode $\v$,
\begin{align}
\v \rightarrow \sqrt{\beta}\,\v, \qquad \text{where } \v \equiv \left( \frac{ a_{n1}+b_{n2}  }{ \sqrt{2} }\right) .
\end{align}
This modifies the measure to
\begin{equation} \label{scaledmeasure}
\frac{ d\delta T}{\frac{E}{\pi n}+\frac{\delta T}{\sqrt{\beta}}}  V\delta\left( x_\mu \right) 
\delta\left( \frac{b_{n1}-a_{n2}}{\sqrt{2}}\right)\left[  \frac{\pi m}{E}+ \sqrt{\beta}\pi
\left(\frac{a_{n1}+b_{n2}}{\sqrt{2}}\right)\right]~\prod_{\mu=1}^4dx_\mu\prod_{k=1}^\infty da_{k\mu}db_{k\mu} .
\end{equation}
Now consider the integral over the mode $\v$, which after the above rescaling becomes
\begin{equation}
	\ldots \int d\v \left(\frac{\pi m}{E}+ \sqrt{\beta} \,\pi \v \right) e^{- \frac{2(\pi n)^2 m}{E} \delta T \left(  \v + \sqrt{\beta} \frac{E}{2m} \v^2\right)} \ldots  .
\end{equation}
The ellipsis stands for the rest of the path integral. In terms of the variable
\begin{equation}
	\xi \equiv \v + \frac{\sqrt{\beta} E}{2m} \v^2 \label{nicolai}
\end{equation}
this is just
\begin{equation}
	\frac{\pi m}{E} \int d\xi \,e^{-\frac{2(\pi n)^2 m}{E} \,\xi\cdot \delta T} ,
\end{equation}
which demonstrates that we can simply drop the terms proportional to $\beta^{+\frac{1}{2}}$ in the measure and action. Equation \eqref{nicolai} is the ``Nicolai map'' that reduces this factor of the path integral to Gaussian form. 

The remaining corrections to the semiclassical approximation, in both the measure and the action, are suppressed by powers of $\sqrt{\beta}$. Now, we remember that the integral is independent of $\beta$.  The original integral that we computed was for the case $\beta=1$.   Assuming smooth behavior in $\beta$, we can set the original integral equal to the limit of the above as $\beta\to \infty$.  In that limit, the interaction terms in the action and in the measure go to zero and the integral is reduced to the Gaussian one which we have already computed in the
previous section where we found that it gives the exact result.

\section{Discussion}

In conclusion, we note that there are circumstances where the world-line path integral in the presence of more general, non-constant electric fields is thought to be exact \cite{Cooper:2006mt}\textsuperscript{,}\cite{Ilderton:2014mla}.  Although we shall not do so here, it would be very interesting to understand whether our results could be extended to those cases.   

One generalization which our results can be considered a preparation for is the inclusion of dynamical gauge 
fields.   That could be done by including the Wilson loop in the word-line path integral, 
\begin{align}
\Gamma ~= ~\frac{1}{V} ~\int_0^\infty\frac{dT}{T }~\int_0^1 [dx^\mu(\tau)]~e^{-\int_0^1d\tau\left[\frac{1}{4T}\dot x^\mu(\tau)^2+\frac{1}{2}F^{\mu\nu}x^\mu(\tau)\dot  x_\nu(\tau)+Tm^2\right]}\left<e^{i\oint d\tau \dot x^\mu(\tau)A_\mu(x(\tau)) }\right>
\end{align}
where the bracket is the expectation value of the operator in the relevant quantum field theory and we have
separated a constant background   field $F^{\mu\nu}$ from the fluctuating gauge field of the quantum field theory. 
The expectation value, $\left<e^{i\oint d\tau \dot x^\mu(\tau)A_\mu(x(\tau)) }\right>$ is a   functional
of the trajectory $x^\mu(\tau)$.  A semi-classical approximation to the amplitude begins with seeking a solution of 
the ``classical'' equation of motion, which now must be derived from the action including the Wilson loop.  
The latter provides a potential whose derivative is a force term which appears in  the equation of motion of the particle
$$
-\frac{1}{2T}\ddot x^\mu(\tau)+F^{\mu\nu}\dot x^\nu(\tau) =\frac{\delta}{\delta x^\mu}\ln\left<e^{i\oint d\tau \dot x^\mu(\tau)A_\mu(x(\tau)) }\right>
$$
By symmetry, in a Euclidean rotation invariant field theory, due to the symmetry of a circle under rotations
about its centre, 
$$
\left. \frac{\delta}{\delta x^\mu(\tau)}\ln\left<e^{i\oint d\tau \dot x^\mu(\tau)A_\mu(x(\tau)) }\right>
\right|_{x^\mu={\rm circle} }=0
$$
In an external electric or magnetic field, knowing that the circle trajectory is still a classical solution, and
the understanding that in the absence of gauge field fluctuations, the semi-classical expansion beginning
wit the circle trajectory leads to the correct result fro the Schwinger formula in the n-instanton section provides
a starting point for studying corrections from quantum fluctuations of the gauge fields.   This idea was first exploited
by Affleck, Alvarez and Manton  \cite{Affleck:1981bma}  to compute the leading correction from photon exchange and it has recently been used to study the strong coupling limit of the Schwinger formula and the behaviour of heavy quarks in electric fields in the context of AdS/CFT holography  \cite{Semenoff:2011ng,Hubeny:2014zna}.

\begin{acknowledgements}
The authors acknowledge financial support of NSERC of Canada. The research leading to these results has received funding from the People Programme (Marie Curie Actions) of the European Union's Seventh Framework Programme FP7/2007-2013/ under REA Grant Agreement No 317089. 
\end{acknowledgements}

\appendix

\section{The Riemann zeta function}  

In this appendix, we review some properties of the zeta function which are needed
in the following appendix to derive the world-line path integral and in Section 2 for the
definition of infinite products and summations which are encountered in the Gaussian functional
integral which is done there.   A more thorough review of zeta functions and relevant discussion
can be found in many references, for example, reference \cite{Elizalde:1994gf} \cite{Elizalde:1996zk}.

 The Riemann zeta function is defined by the infinite sum
\begin{equation}
\zeta(s)= \sum_{k=1}^\infty \frac{1}{k^s}
\label{zeta}
\end{equation}
defined as a function of a complex variable $s$ where the real part of $s$ should
be large enough so that the sum converges.   The function is then analytically
continued to the entire complex plane where it is a meromorphic function on the 
whole complex $s$-plane, which is holomorphic everywhere except for a simple pole at $s = 1$, 
with residue 1.

The values of the zeta function and its derivative which we use are
\begin{equation}\label{zetazero}
\zeta(0)=\lim_{s\to0} \sum_{k=1}^\infty \frac{1}{k^s}=-\frac{1}{2}~\to~
\prod_{k=1}^\infty \alpha =\lim_{s\to 0} \prod_{k=1}^\infty \alpha^{\ \frac{1}{k^s}}= \alpha^{\zeta(0)}=\alpha^{-\frac{1}{2}}
\end{equation}
and
\begin{align}
\zeta'(0)&=\lim_{s\to 0}\frac{d}{ds}   \sum_{k=1}^\infty \frac{1}{k^s}=-\lim_{s\to0} \sum_{k=1}^\infty \frac{\ln k}{k^s}  
=-\frac{1}{2}\ln 2\pi~\\  
\prod_{k=1}^\infty k &= \lim_{s\to 0}\prod_{k=1}^\infty e^{\frac{\ln k}{k^s}} = e^{-\zeta'(0)}=(2\pi)^{\frac{1}{2}} 
\label{zetaprimezero} 
\end{align}
A consequences of (\ref{zetaprimezero})  which we shall use is
\begin{equation}\label{simpleidentity2}
\prod_{k=1}^\infty (2\pi k) = 1
\end{equation}
 
\section{World line path integral}

In this appendix, we will demonstrate that the one-loop vacuum energy of a scalar particle is given exactly
by the world-line path integral when we define the various infinite products and sums which occur in the latter
using zeta function regularization. The result will be equation (\ref{world-linepathintegral}).

We begin with the usual expression for the vacuum energy density of a complex scalar field in Euclidean space
\begin{align}
\Gamma = \int \frac{d^4p}{(2\pi)^4}\ln(p^2+m^2)
\end{align}
which is represented by the Feynman diagram in figure \ref{vacuumenergy}.  Recall that a real scalar field would
be the same expression with a factor of 1/2 in front.  At this level, a complex scalar field is simply two real scalar
fields which have twice as much vacuum energy. 
{\begin{figure}
\includegraphics[scale=0.6]{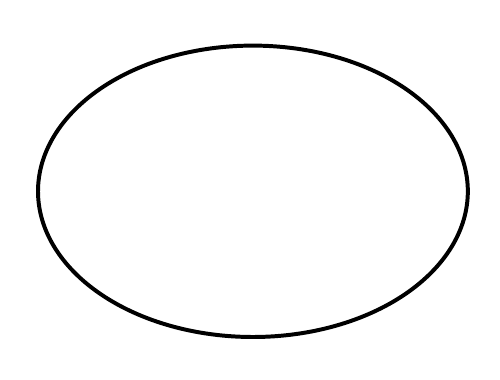}~~~
\begin{caption} { The Feynman diagram which must be computed to find the vacuum
energy of a scalar field.
  \label{vacuumenergy}
}\end{caption}
\end{figure}
Here, the four-momentum is Euclidean and a space-time volume factor has been removed so that
the result of doing the integral is the energy density. We will reorganize this integral
in order to represent it as a world-line path integral.  The singular integrals which we encounter
will de defined by zeta function regularization.  To proceed, we first introduce a Schwinger
parameter,  $T$, 
\begin{equation}\label{partitionfunction}
\Gamma =  \int_0^\infty\frac{dT}{T }~\int\frac{d^4p}{(2\pi)^4}
e^{-T(p^2+m^2)}
\end{equation}
We note that, in four dimensions, this integral contains an ultraviolet divergence, coming
from the $T\sim 0$ integration region.  This divergence must be regulated in order to obtain
a sensible definition of the vacuum energy.  We could regulate this expression by defining it
as the limit
\begin{align}
\ln(p^2+m^2)=\lim_{\kappa\to0}\frac{d}{d\kappa}(p^2+m^2)^\kappa
=\lim_{\kappa\to 0} \frac{d}{d\kappa} \frac{1}{\Gamma[\kappa]}
\int \frac{dT}{T^{1+\kappa}}e^{-T(p^2+m^2)}
\end{align}
The appropriate quantity to study would then be $\int \frac{dT}{T^{1+\kappa}}e^{-T(p^2+m^2)}$ with the exponent
of $T$ shifted by $\kappa$ which we could always take as negative with large enough magnitude that the integrals
to be done converge, and then define the quantity in the region near $\kappa=0$ by analytic continuation. In the
following, we shall assume that this regulator is implicitly there, if needed but we will stick with expression (\ref{partitionfunction}) 
as the writing will be slightly simpler. 

Now,  consider the functional integral 
\begin{align}
\int [dx^\mu(\tau)]~e^{i\int_0^1 d\tau p_\mu(\tau) \tfrac{d}{d\tau} x^\mu(\tau)}
\label{functionalintegral1}\end{align}
where both $x(\tau),p(\tau)$ have periodic boundary conditions, $x^\mu(\tau+1)=x^\mu(\tau)$ and
$p_\mu(\tau+1)=p_\mu(\tau)$.   

It is very convenient to 
use the expansions of the integration
variables in a discrete orthonormal 
complete set of periodic functions, 
\begin{align}
p_\mu(\tau) = p_\mu+\sum_{k=1}^\infty \left[p_{\mu k}\sqrt{2}\sin(2\pi k\tau)+\tilde p_{\mu k}\sqrt{2}\cos(2\pi k\tau)\right]
\end{align}
and
\begin{align}
x^\mu (\tau) = x^\mu+\sum_{k=1}^\infty \left[x^\mu_k\sqrt{2}\sin(2\pi k\tau)+
\tilde x^\mu_k\sqrt{2}\cos(2\pi k\tau)\right]
\end{align}
The  complete set of orthonormal periodic functions is $(1,\sqrt{2}\sin(2\pi k\tau), \sqrt{2}
\cos(2\pi k\tau) )$, where the normalization is the square-integral over the interval $\tau\in[0,1]$. 
The functional integration measure is then defined as the ordinary Riemann integral over each of
an infinite number of real variables, 
\begin{align}
[d x^\mu(\tau)]\equiv dx^\mu\prod_{k=1}^\infty dx_k^\mu d\tilde x_k^\mu ~~,~~
[d p_\mu(\tau)]\equiv dp_\mu\prod_{k=1}^\infty dp_{\mu k} d\tilde p_{\mu k} 
\end{align}
Now, with these definitions, consider 
\begin{align}
&\int [dx^\mu(\tau)]\exp(i\int_0^1 d\tau p_\mu(\tau)\frac{d}{d\tau}  x^\mu(\tau)) 
\nonumber
\\
&
=\int dx^\mu \prod_{k=1}^\infty dx^\mu_k d\tilde x^\mu_k \exp\left(i
 \sum_{k=1}^\infty (2\pi k ) [p_{\mu k}\tilde x^\mu_k-\tilde p_{\mu k}x^\mu_k]\right)
 \nonumber \\
& = V\prod_{\mu=1}^4\prod_{k=1}^\infty (2\pi)\delta( (2\pi  k) p_{\mu k})\cdot (2\pi)\delta( (2\pi  k) \tilde p_{\mu k})
\nonumber \\
\nonumber 
&
= V  \left(\frac{1}{\prod_1^\infty k }\right)^8\prod_{k= 1}^\infty \delta( p_{\mu k})\delta( \tilde p_{\mu k})
= V  \left(\frac{1}{\exp(-\zeta'(s))}\right)^8\prod_{k= 1}^\infty \delta( p_{\mu k})\delta( \tilde p_{\mu k})
\label{36} \\ 
&=V\frac{1}{(2\pi)^4}\prod_{k=1}^\infty \delta( p_{\mu k})\delta( \tilde p_{\mu k})
\nonumber
\end{align}
where we have used
$$
\lim_{s\to0}~~\prod_1^\infty  k   = \lim_{s\to0}~\exp\left(-\frac{d}{ds}\sum_{k=1}^\infty k^{-s}\right)=e^{-\zeta'(0)} 
$$
and  $V\equiv \int dx^\mu$ is the (infinite) space-time volume arising from the integral over the constant mode $x^\mu$
and the fact that it does not appear in the integrand.   

 The identity          
\begin{align}
 \int [dx^\mu(\tau)]\exp(i\int_0^1 d\tau p_\mu(\tau)  \frac{d}{d\tau} x^\mu(\tau)) 
=V\frac{1}{(2\pi)^4}\prod_{k=1}^\infty \delta( p_{\mu k})\delta( \tilde p_{\mu k})
\end{align}
tells us that, if we first do the integral over $x^\mu(\tau)$ in the following path functional integral
\begin{align}
\int \frac{d^4p}{(2\pi)^4} e^{-T(p^2+m^2)}=
 \frac{1}{V}\int [dx^\mu(\tau)][dp_\mu(\tau)]~ e^{\int_0^1\left[i p_\mu(\tau)\tfrac{d}{d\tau} x^\mu(\tau)
-T(p_\mu(\tau) p_\mu(\tau)+m^2)\right]}
\label{identity}
\end{align}
it will generate a factor of $V$ and delta functions for all of the nonzero modes of $p_\mu(\tau)$.  The integrals over those nonzero modes
can then be done, leaving the integral over the constant mode in $p_\mu(\tau)$ which becomes the momentum that appears on the left-hand-side of the equation.  
Now, we reorganize the right-hand-side of (\ref{identity}) by doing the functional integral over $p_\mu(\tau)$.

\begin{align}
\int &[dp_\mu(\tau)]~ e^{\int_0^1\left[i p_\mu(\tau)\tfrac{d}{d\tau} x^\mu(\tau)-T(p_\mu^2(\tau)+m^2)\right]}
\nonumber \\
&=\int dp_\mu \prod_{k=1}^\infty dp_{\mu k}d\tilde p_{\nu k}~ e^{\sum_{k=1}^\infty\left[(2\pi ki) [p_{\mu k}\tilde x^\mu_k
-\tilde p_{\mu k}x^\mu_k]-T(p_{\mu k}^2 +\tilde p_{\mu k}^2)\right]-Tp^2}
\nonumber \\
&=\sqrt{\frac{\pi}{T}\left(\prod_{k=1}^\infty \frac{\pi}{T}\right)^2 }e^{-\frac{1}{4T}\sum_{k=1}^\infty\left[(2\pi k  x^\mu_k)^2+(2\pi k\tilde x^\mu_k)^2\right]}
=e^{-\frac{1}{4T}\int_0^1d\tau \left(\frac{d}{d\tau}x^\mu(\tau)\right)^2 } \label{37}
\end{align}
where we have used the fact that the pre-factor is $\sqrt{\frac{\pi}{T}\left(\prod_{k=1}^\infty \frac{\pi}{T}\right)^2 }=\left({\frac{\pi}{T} }\right)^{\zeta(0)+1/2}=1$.
The result is then
\begin{align}
\int \frac{d^4p}{(2\pi)^4} e^{-T(p^2+m^2)}=\frac{1}{V}\int [dx^\mu(\tau)]  e^{\int_0^1\left[-\frac{1}{4T}\dot x^\mu(\tau)^2-Tm^2\right]}
\end{align}
 where the dot denotes $\tau$-derivative and we have the path integral formula for the vacuum energy density of a complex scalar field
\begin{equation}
\boxed{
\Gamma ~= ~\frac{1}{V} ~\int_0^\infty\frac{dT}{T }~\int [dx^\mu(\tau)]~e^{-\int_0^1d\tau \left[\frac{1}{4T}\dot x^\mu(\tau)^2+Tm^2\right]}
~,~x_\mu(\tau+1)=x_\mu(\tau)}
\label{world-linepathintegral}
\end{equation}
What we have shown is that, if one uses zeta function regularization,
equation (\ref{world-linepathintegral}) is an identity.  

We can check this identity by doing the path integral directly.  
We begin by doing the functional integral in the world-line expression (\ref{world-linepathintegral})
that we
 have derived.  The integral over the constant mode of $x^\mu(\tau)$ gives a volume factor which
cancels the factor of $1/V$.
 We can then use the rules for doing Gaussian integrals to 
do the quadratic functional integral over nonzero modes of $x^\mu(\tau)$.  The
 result is
 \begin{align}
\Gamma ~= ~ \int_0^\infty\frac{dT}{T }~ e^{ -Tm^2 }\left[\prod_{k=1}^\infty\frac{2\pi\cdot 2T}{ (2\pi k)^2}\right]^{4}
= ~ \int_0^\infty\frac{dT}{T }~ e^{ -Tm^2 }\left[ 4\pi T \right]^{4\zeta(0)}
\nonumber \\
= ~ \int_0^\infty\frac{dT}{T }~ e^{ -Tm^2 }\frac{1}{ (4\pi T)^2}
\label{98}\end{align}
Again, we have used zeta function regularization to define the infinite products. 
The result is identical to what is obtained by integrating (\ref{partitionfunction}) over $p$.

Note that the variable $T$ here is the inverse
of the one that we use in the body of this paper (and can simply be gotten by performing the 
change of variable $T\to 1/T$).

Now, consider the vacuum energy of a charged scalar particle coupled to an electromagnetic field whose
vector potential is $A_\mu(x)$.  The coupling is implemented in the Euclidean functional integral by
including the Bohm-Aharonov phase factor, to obtain 
\begin{align}
\Gamma ~= ~\frac{1}{V} ~\int_0^\infty\frac{dT}{T }~\int_0^1 [dx^\mu(\tau)]~e^{-\int_0^1d\tau\left[\frac{1}{4T}\dot x^\mu(\tau)^2+Tm^2\right]+i\oint d\tau \dot x^\mu(\tau)A_\mu(x(\tau)) }
\end{align}
In particular, if we couple to an external constant electric field $E$, the gauge field could be taken as
\begin{align}
A_\mu(x)=(0,iEx^1(\tau),0,0)
\end{align}
A physical electric field obtains a factor of $i$ in Euclidean space.
Now the path integral would be the Gaussian integral 
\begin{align}\label{vacuumenergywithelectricfield}
\Gamma ~= ~\frac{1}{V} ~\int_0^\infty\frac{dT}{T }~\int [dx^\mu(\tau)]~e^{-\int_0^1 \left[\frac{1}{4T}\dot x^\mu(\tau)\dot x^\mu(\tau)+E
x^1(\tau)\dot x^2(\tau) + Tm^2\right]
} 
\end{align}
This integral should yield the vacuum energy of a charged scalar field which is coupled to an external electric field.  
 
 \subsection{A note on regularization}
 
 The zeta function regularization that we have done can be performed by systematically altering the 
 functional integral so that the infinite products that are generated when the Gaussian integrals are
 performed give zeta-functions.  In order to avoid cluttering the equations with notation and limits, we have omitted discussion of this in the text, however it is worthy of 
 comment here.  We can modify the quadratic form in the exponent of the functional
 integral so that
 $$
- \frac{1}{4T} \frac{d^2}{d\tau^2} \delta(\tau-\tau') = \sum_{k\in{\mathcal Z}} e^{2\pi ik(\tau-\tau')}\frac{ (2\pi k )^2}{4T}
$$
is replaced by the non-local operator 
$$
 O_s(\tau-\tau') = \pi \sum_{k\in{\mathcal Z}} e^{2\pi ik(\tau-\tau')} \left[\frac{(2\pi k )^2}{4\pi T}\right]^{|k|^{-s}}
$$
 We see that 
 $$
\lim_{s\to 0}  O_s(\tau-\tau') =- \frac{1}{4T} \frac{d^2}{d\tau^2} \delta(\tau-\tau') 
$$
  Also, 
  $$
    {\det}' \frac{1}{\pi} O_s = \prod_{k=0}^\infty \left[\frac{(2\pi k )^2}{4\pi T}\right]^{2|k|^{-s}}
  = e^{2\zeta(s) \ln\left[\frac{\pi }{ T}\right]  -4\zeta'(s) }
  $$
  where $\det'$ means that we leave the zero mode out of the determinant.

  \begin{align}
\Gamma ~= \lim_{s\to 0}~\frac{1}{V} ~\int_0^\infty\frac{dT}{T }~\int [dx^\mu(\tau)]~e^{-\int_0^1d\tau \left[x^\mu{\mathcal O}_sx^\mu(\tau)+Tm^2\right]}  
\nonumber \\
=\lim_{s\to 0} ~ \int_0^\infty\frac{dT}{T }~ e^{ -Tm^2 }
\left[\frac{\pi }{ T}\right]^{-D\zeta(s)  }e^{ 2D\zeta'(s) }
\end{align}
 where $D$ is the space-time dimension. This
  reproduces equation (\ref{98}) when $D=4$.

  When the electric field is applied, in the $n$-instanton sector, it is
  convenient to begin with gauge-fixed functional integral 
  \begin{align}\label{integral_to_be_done}
\int \prod_{\mu=1}^4dx_\mu\prod_{k=1}^\infty da_{k\mu}db_{k\mu}\frac{ d\delta T}{\frac{E}{\pi n}+\frac{\delta T}{\sqrt{\beta}}}  V\delta\left( x_\mu \right) 
\delta\left( \frac{b_{n1}-a_{n2}}{\sqrt{2}}\right)\left[  \frac{\pi m}{E}+ \frac{\pi}{\sqrt{\beta}}
\left(\frac{a_{n1}+b_{n2}}{\sqrt{2}}\right)\right]e^{-S}
\end{align}
with the action
\begin{align}
S=& \frac{\pi n m^2}{E}
+ \frac{2m^2(\pi n)^3 }{\beta E^3} \frac{\delta T^2}{2} 
\nonumber \\ 
&+ \frac{ m(2\pi n)^2}{2   E}  \delta T  \left( \frac{ a_{n1}+b_{n2}  }{ \sqrt{2} }\right) 
+\frac{4\pi nE}{2 }\left(\frac {a_{n1}-b_{n2} }{\sqrt{2}}\right)^2 +
\frac{4\pi nE}{2  }\left(\frac{a_{n2}+b_{n1}}{\sqrt{2}}\right)^2        \nonumber \\ 
&+  \frac{E}{4\pi n  }\sum_{k=1,\neq n}^\infty (2\pi k)^2\left[(a_{k\mu}^2+b_{k\mu}^2)-\frac{2n}{k}(a_{k1}b_{k2}-a_{k2}b_{k1})\right]\nonumber \\
&+\frac{1}{4\beta^{\frac{1}{2}}}\delta T\sum_{k=1}^\infty(2\pi k)^2[(a_k^\mu)^2+(b_k^\mu)^2]+\sum_{k=3}^\infty m^2\frac{1}{\beta^{\frac{k}{2}}}
\left(\frac{\pi n}{ E}\right)^{k+1}(-\delta T)^k
\end{align}
We regulate this integral by making the following two  replacements in the infinite sums in the action 
$$
 \frac{E}{4\pi n  }\sum_{k=1,\neq n}^\infty (2\pi k)^2 \ldots \to \pi \sum_{k=1,\neq n}^\infty \left[\frac{E}{4\pi^2 n  }(4\pi k)^2
 \right]^{k^{-s}}\dots
 $$
 and 
 $$
 \frac{1}{4\beta^{\frac{1}{2}}}\delta T\sum_{k=1}^\infty(2\pi k)^2\ldots\to
 \frac{1}{4\beta^{\frac{1}{2}}}\delta T\sum_{k=1}^\infty[(2\pi k)^2]^{k^{-\tilde s}}\ldots 
 $$
 where $s$ and $\tilde s$ have sufficiently large real parts.  We will then define the quantities which we
 compute for all values of the complex variables $s$ and $\tilde s$.  We are interested in taking the limits
 as $s\to 1$ and $\tilde s\to 1$ of those functions. We find that the relevant quantities have finite limits. 
 For example,  the determinant of the quadratic form contains the product 
  $$\prod_{k\neq 0} \left[\frac{E}{ 4\pi^2 n}(2\pi k)^2\right]^{2k^{-s}}
 =\exp\left( 4\zeta(s)\ln\left[\frac{4\pi^2 E}{n}\right]-4\zeta'(s) \right)
    $$
    which is finite for all $s\neq 1$.  The regularization of the interaction turns out to be sufficient to make
    all of the contributions, order by order in perturbation theory, finite, if the real part of $s$ is sufficiently large.
    As an example, let us compute the leading correction in an asymptotic expansion of the integral
    in (\ref{integral_to_be_done}) in $\frac{1}{\sqrt{\beta}}$.  This can be done using the standard Dyson-Wick technique
    which begins with the leading order two-point correlation functions of the variables. 
    The non-vanishing two-point correlation functions are 
    \begin{align}\nonumber 
    &\left<\delta T \left( \frac{b_{n1}-a_{n2}}{\sqrt{2}}\right) \right>_0 =  \frac{ 2 E}{ m(2\pi n)^2}
    \\ \nonumber 
    &\left<a_k^\mu a_{k'}^\nu\right>_0=\delta_{kk'}\delta^{\mu\nu}
    \frac{1}{2\pi}\left[\frac{E}{4\pi^2 n  }(4\pi k)^2
 \right]^{-k^{-s}}
=\left<b_k^\mu b_{k'}^\nu\right>_0~,~k\neq 0,n
\\ \nonumber &
    \left<\left(\frac{ a_{n1}+b_{n2}  }{ \sqrt{2} }\right)\left(\frac{ a_{n1}+b_{n2}  }{ \sqrt{2} }\right)\right>_0 = \frac{1}{4\pi n E}
    \\ \nonumber &
    \left<\left(\frac{a_{n1}-b_{n2} }{\sqrt{2}}\right)\left(\frac {a_{n1}-b_{n2} }{\sqrt{2}}\right)\right>_0=  \frac{1}{4\pi n E}
    \end{align}
    
 The term, which would be of order $\frac{1}{\beta^{\frac{1}{2}}}$,  vanishes
    by symmetry.   The next-to-leading term is of order $\frac{1}{\beta}$.  It is 
    given by Wick contractions of the correlation function
   $$ \frac{1}{\beta}\frac{\pi n}{E}\frac{\pi m}{E} \left< \frac{\pi n}{E}\delta T~\frac{\pi n}{E}\delta T~
    -\frac{\pi n}{E}\delta T~\frac{E}{m}\left(\frac{a_{n1}+b_{n2}}{\sqrt{2}}\right) 
    +\frac{\pi n}{E}\delta T~\frac{1}{4}\delta T\sum_{k=1}^\infty(2\pi k)^{2k^{-\tilde s}}[(a_k^\mu)^2+(b_k^\mu)^2]
  \right. 
  $$
  $$
  \left.   
    -\frac{E}{m}\left(\frac{a_{n1}+b_{n2}}{\sqrt{2}}\right) ~ \frac{1}{4}\delta T\sum_{k=1}^\infty(2\pi k)^{2k^{-\tilde s}}
    [(a_k^\mu)^2+(b_k^\mu)^2]
 \right.
  $$
  $$
  \left.
     + \frac{1}{2!}\frac{\delta T}{4}\sum_{k=1}^\infty(2\pi k)^{2k^{-\tilde s}}[(a_k^\mu)^2+(b_k^\mu)^2] ~~\frac{\delta T}{4} \sum_{\tilde k=1}^\infty(2\pi \tilde k)^{2\tilde k^{-\tilde s}}[(a_{\tilde k}^\mu)^2+(b_{\tilde k}^\mu)^2]\right>_0
  $$
  which becomes 
 $$ \frac{1}{\beta}\frac{\pi n}{E}\frac{\pi m}{E} \left\{ ~0~
    - ~\frac{\pi n}{E}\frac{E}{m}  \frac{  2E}{ m(2\pi n)^2}   
  ~ +~\frac{\pi n}{E}\frac{1}{2} (2\pi n)^{2n^{-\tilde s}} \left(\frac{  2E}{ m(2\pi n)^2} \right)^2
  \right. 
  $$
  $$
  \left.   
    -\frac{E}{m} ~ \frac{1}{4} \frac{ 2 E}{ m(2\pi n)^2} \left<\sum_{k=1}^\infty(2\pi k)^{2k^{-\tilde s}} [(a_k^\mu)^2+(b_k^\mu)^2]\right>_0
 \right.
  $$
  $$
   \left.
     + \frac{1}{2!}\frac{1}{4}(2\pi n)^{2n^{-\tilde s}}
     \left(\frac{  2E}{ m(2\pi n)^2} \right)^2\left<\sum_{k=1}^\infty(2\pi k)^{2k^{-\tilde s}}[(a_k^\mu)^2+(b_k^\mu)^2] \right>_0 \right\}
  $$
  We can easily
  see that, when the regulator is removed, the terms cancel identically, as we expected.  To find the cancelation, 
  we do not need to evaluate the tadpoles, $ \left<\sum_{k=1}^\infty(2\pi k)^{2k^{-\tilde s}} [(a_k^\mu)^2+(b_k^\mu)^2]\right>_0
$.    However, the zeta-function regularization does render them finite.

\section{A simple example}
 
 In this appendix, we shall test the saddle point approximation for the case of an  integral which is similar  to, 
but is much simpler than the world-line path integral
(\ref{vacuumenergywithelectricfield}) that we discussed in the Appendix above. The idea is to compute
\begin{align}
I=\Im \int_0^\infty\frac{dT}{T}\int d^2x ~e^{-\frac{m^2}{T} - \left(T-T_0\right)\vec x^2}
\end{align}
The integral over $\vec x$ is well-defined when $T>T_0$ and it diverges when $T<T_0$.  
We shall define the integral by assuming that $T$ is in the region where the $\vec x$-integration is well-defined,
doing the integral and continuing the result to the entire $T$-plane.  What we then find is a pole at $T=T_0$. 
This analytic continuation results in the integral having an imaginary part due to the pole, where the integration
contour must be defined using an ``$i\epsilon$ prescription''.  
The gaussian integral
over $\vec x$ produces
\begin{align}
I=\Im \pi \int_0^\infty\frac{dT}{T} ~e^{-\frac{m^2}{T} }\frac{1}{T-T_0}
\end{align}
The prescription is to replace this integral by 
\begin{align}
I=\Im \pi \int_0^\infty\frac{dT}{T} ~e^{-\frac{m^2}{T} }\frac{1}{T-T_0+i\epsilon}
\end{align}
and to use the formula
\begin{align}
\frac{1}{T-T_0+i\epsilon}=\frac{\cal P}{T-T_0}-i\pi\delta(T-T_0)
\end{align}
so that
\begin{align}
I=- \frac{\pi^2}{T_0} ~e^{-\frac{m^2}{T_0} } 
\end{align}
Now, let us do the integral using the  saddle point technique.  We consider the 
action 
\begin{align}
S=\frac{m^2}{T} + \left(T-T_0\right)\vec x^2
\end{align}
The classical equations of motion are
\begin{align}
\frac{m^2}{T^2}= \vec x^2
~~,~~
\left(T-T_0\right)\vec x=0
\end{align}
These have a solution where 
\begin{align}
\vec x=\frac{m}{T_0} \hat n
~~,~~ T=T_0 
\end{align}
with $\hat n$  an arbitrary unit vector. 
The action evaluated on this solution is 
\begin{align}
S_{\rm cl}=\frac{m^2}{T_0}
\end{align}
Then, we consider fluctuations, 
\begin{align}
\vec x=\frac{m}{T_0}\hat n+\delta \vec x
~~,~~ T=T_0 +\delta T
\end{align}
The quadratic approximation to the action is 
\begin{align}
S= \frac{m^2}{T_0}+\frac{2m^2}{T_0^3 }\frac{\delta T^2}{2}
+ \frac{2m}{T_0}\delta T\hat n\cdot\delta x 
+\ldots
\end{align}
The Gaussian integral over $\delta T$ produces a measure factor
\begin{align}
\sqrt{\frac{\pi T_0^3}{m^2}}
\end{align}
and the remaining action becomes
\begin{align}
S_1= \frac{m^2}{T_0}  -2T_0\frac{(\hat n\cdot\delta\vec x)^2}{2}
+\ldots
\end{align}

Now, the degree of freedom $\hat n\cdot \delta x$ is
tachyonic and its integral produces the measure factor
\begin{align}
\sqrt{-\frac{\pi}{ T_0}}
\end{align}
As well, there is an integral over the other component, $\hat n\times \delta\vec x$
which appears to be divergent.  This apparent divergence is due to a symmetry which must
be handled by the collective coordinate technique.  For this purpose, we introduce the
identity
\begin{align}
1=\frac{1}{2}\int_0^{2\pi} d\theta \delta (\hat n_\theta\times\vec  x) \left|\frac{d}{d\theta}\left[\hat n_\theta\times\vec  x\right]\right|
\end{align}
into the original integral.\footnote{The factor of 2 cancels a Gribov copy. 
If $\hat n=(1,0)$, then $n_\theta = (\cos\theta, \sin\theta)$ and if $\vec x=x(\cos\phi,\sin\phi)$, 
$\hat n_\theta\times\vec x=\sin(\theta-\phi) x$ which has two zeros in the range $0\leq\theta<2\pi$. The factor of $1/2$ cancels
this multiplicity of zeros of the constraint. }
Then, by changing the integration variable in the integrand $\delta x\to \delta x_\theta$ the entire integral becomes independent
of $\theta$. 

The Jacobian becomes
\begin{align}
\left.\frac{d}{d\theta} \hat n_\theta\times\vec x_{\theta'}\right|_{\theta'=\theta}=\hat n\cdot (\frac{m}{T_0}\hat n+\delta x)
=\frac{m}{T_0}+\ldots
\end{align}
and the integration over $\theta$ produces an additional factor of $2\pi$.  Gathering the measure factors, we find the result
\begin{align}
 I = \Im \frac{e^{-\frac{m^2}{T_0}}}{T_0}\cdot \sqrt{\frac{\pi T_0^3}{m^2}}  \cdot \sqrt{\frac{-\pi}{ T_0}}\cdot \frac{1}{2}\cdot \frac{m}{T_0}
\cdot 2\pi 
=\pm ~\frac{e^{-\frac{m^2}{T_0}}}{T_0}\cdot \pi^2
\end{align}
This produces the result of the first, exact evaluation but with a sign ambiguity. The correct sign must be chosen to reproduce the exact
result. 

We have found the exact result for the imaginary part of the integral by computing the integral in the ``instanton'' sector
to the leading and next-to-leading order in a saddle point approximation.   This suggests that the higher order corrections
must all vanish. 

As we will not show, it is straightforward to prove this.
Let us begin with the gauge-fixed integral which we want to compute
\begin{align}
I=\Im \int_0^\infty\frac{dT}{T}\int d^2x ~e^{-m^2/T - \left(T-T_0\right)\vec x^2}\frac{1}{2}\int_0^{2\pi} d\theta \delta (\hat n_\theta\times\vec  x) 
 \left|\hat n_\theta\cdot\vec  x\right|
 \end{align}
 This expression is identical to the original integral, plus we have inserted the Fadeev-Popov identity. We ``gauge fix'' the integral
 by transforming the variables $\vec x$ by a rotation by angle $\theta$.  Then, the entire integrand is independent of $\theta$
 and we can do the $\theta$-intergal.   The result is 
 \begin{align}
I=\Im \int_0^\infty\frac{dT}{T}\int d^2x ~e^{-m^2/T - \left(T-T_0\right)\vec x^2}\pi\delta (\hat n\times\vec  x) 
 \left|\hat n\cdot\vec  x\right|
 \end{align}
 To proceed, we perform the change of variables, 
$\vec  x = \alpha \vec {\tilde x}$, $  T=\tilde T/\alpha^2$, with $\alpha$ a positive real number.  The integral
becomes (dropping the tildes after we have changed the variables)
\begin{align}
I=\Im \int_0^\infty\frac{dT}{T}\int d^2x ~\alpha^2~e^{-\alpha^2 m^2/T - \left(T-\alpha^2 T_0\right)\vec x^2}
\pi\delta (\hat n\times\vec  x) \left|\hat n\cdot\vec  x\right|
\end{align}
Note that the factors of $\alpha$ cancel from the last two terms.  All we have done here is a change of the 
integration variable.   That change is parameterized by $\alpha$.  The final integral cannot depend on $\alpha$.
We will take advantage of this fact shortly.

Now, let us study the semiclassical expansion of the integral.   The classical equations of motion are
$$
\alpha^2\frac{m^2}{T^2}=\vec x^2~~,~~
\left(T-\alpha^2 T_0\right)\vec x=0
$$
They are solved by $
T=\alpha^2 T_0$ and $\vec x=\hat n \frac{m}{T_0}$
and we make the substitutions
$$
T=\alpha^2 T_0+\tau~~,~~\vec x=\hat n \frac{m}{T_0}+\vec y
$$
The integration measure is $dTd^2x=[d\tau][d(\hat n\cdot\vec y)][d(\hat n\times\vec y)]$ and the action becomes 
$$
S 
=\frac{m^2}{T_0} + 2\frac{m}{T_0}~\tau \hat n\cdot \vec y + \frac{m^2}{\alpha^2T_0^3}\tau^2+\frac{m^2}{T_0} \sum_{k=3}^\infty
\frac{(-\tau)^k}{\alpha^{2k}T_0^k} + \tau (\hat n\cdot\vec y)^2+\tau(\hat n\times\vec y)^2
$$
Now, we change variables again,  $\tau= {\alpha}\tilde \tau$ and $\vec y= \vec {\tilde y}/{\alpha}$, so that the quadratic terms in the
action become $\alpha$-independent.  With this change, and dropping the tildes after the change of variables is completed, 
the  integral is 
\begin{align}
&I=\Im \int_0^\infty\frac{[d\tau ][d\hat n\cdot \vec y][ d\hat n\times \vec y]}{T_0+\frac{ \tau}{\alpha}}~\pi \delta (\hat n \times\vec  y) \left|\frac{m}{T_0}+
\frac{1}{\alpha}\hat n \cdot\vec  y \right| \cdot
\nonumber \\
&\cdot \exp\left(-\left[\frac{m^2}{T_0} + 2\frac{m}{T_0}~\tau \hat n\cdot \vec y + \frac{m^2}{ T_0^3}\tau^2+\frac{m^2}{T_0} \sum_{k=3}^\infty(-1)^k\frac{\tau^k}{\alpha^{k}T_0^k} +\frac{1}{\alpha} \tau (\hat n\cdot\vec y)^2\right]\right)
\nonumber \end{align}
As we noted above, the integral must be independent of $\alpha$. We are this free to adjust $\alpha$ and to take the limit $\alpha\to\infty$.
In that limit, the integral becomes
\begin{align}
&I=\Im \int_0^\infty\frac{[d\tau ][d\hat n\cdot \vec y][ d\hat n\times \vec y]}{T_0}\delta (\hat n \times\vec  y) \frac{\pi m}{T_0}
\exp\left(-\left[\frac{m^2}{T_0} + 2\frac{m}{T_0}~\tau \hat n\cdot \vec y + \frac{m^2}{ T_0^3}\tau^2\right]\right)
\nonumber
\end{align}
which is identical to the Gaussian approximation to the original integral. This shows that the corrections to the leading and next-to-leading
terms in the semi-classical expansion of the integral give the exact answer.  In the above, we have already confirmed this by explicit calculation. 

\section{Proof without scaling: order-by-order cancellations}
As an alternative to the scaling/change-of-variables argument of section \ref{exactness} we now present a perturbative proof of cancellations of all corrections. First note that all higher-order terms in the action, which we collectively denote $S_{int}$, as well as corrections to the factor $\frac{1}{T}\approx\frac{1}{T_0}$ in the measure, are proportional to $\delta T^p$, ($p\in \mathbb{Z}^+$). If we first separate out the factor $e^{-(\pi n \v)^2\delta T}$, and then Taylor expand the remaining $e^{-\tilde{S}_{int}}$, the path integral becomes a sum of expectation values of monomials in $\delta T$. Focusing on the $\v$, $\delta T$ part
\begin{eqnarray}
	\int d \v \, d \delta T \left( 1+ \frac{E}{m} \v\right) e^{-\frac{A}{2} \delta T^2 -B \v \delta T } e^{-(\pi n \v)^2 \delta T} \, \delta T^p &\equiv& \left\langle ( 1+ \frac{E}{m} \v)e^{-(\pi n \v)^2 \delta T} \delta T^p \right\rangle\\
    &=& \sum_{k=0}^\infty \frac{ (-1)^k (\pi n)^{2k}}{k!} \left(\left\langle \v^{2k} \delta T^{k+p} \right\rangle +\frac{E}{m} \left\langle \v^{2k+1} \delta T^{k+p}\right\rangle \right)
\end{eqnarray}
Performing the $\delta T$ integration (and analytically continuing $\v\rightarrow i\v$ as in section \ref{semiclassical}) one obtains
\begin{equation} \label{sum}
	i\sqrt{\frac{2\pi}{A}} \int d\v \left[ \sum_{k=p}^\infty \frac{(-1)^{k} (\pi n)^{2k}}{B^{k+p}k!} \frac{(2k)!}{(k-p)!} (i \v)^{k-p} + \sum_{k=p-1}^\infty \frac{(-1)^{k} (\pi n)^{2k}}{B^{k+p}k!} \frac{(2k+1)!}{(k-p+1)!} \frac{E}{m} (i \v)^{k-p+1} \right] e^{-(B \v)^2/2A}
\end{equation}
Now since $B=\frac{2(\pi n)^2 m}{E}$, there is an exact term-by-term cancellation between these two sums when $p>0$. For $p=0$ the only difference is that the first term in the right-hand sum is absent, and therefore the first term in the left-hand sum is not cancelled. This term gives precisely the leading, semi-classical contribution to the path integral.

\end{document}